\documentclass[twocolumn,showpacs,preprintnumbers,amsmath,amssymb,a4paper]{iopart} 

\expandafter\let\csname equation*\endcsname\relax
  \expandafter\let\csname endequation*\endcsname\relax

\usepackage{amsmath}
\usepackage[numbers]{natbib}
\usepackage{bbm}
\usepackage{graphicx}
\usepackage{amssymb}
\usepackage[colorlinks=true,pdfstartview=FitH,linkcolor=blue,citecolor=blue,urlcolor=blue]{hyperref}

\usepackage[table]{xcolor} 
\newcommand{\be}{\begin{equation}}
\newcommand{\ee}{\end{equation}}
\newcommand{\ba}{\begin{array}}
\newcommand{\ea}{\end{array}}
\newcommand{\bqa}{\begin{eqnarray}}
\newcommand{\eqa}{\end{eqnarray}}

\newcommand{\um}{\mathbbm{1}}

\newcommand{\bra}[1]{\ensuremath{\langle #1 |}}
\newcommand{\ket}[1]{\ensuremath{| #1 \rangle}}
\newcommand{\prj}[1]{\ensuremath{| #1 \rangle \langle #1 |}}
\newcommand{\ovl}[2]{\ensuremath{\langle #1 | #2 \rangle}}
\newcommand{\matel}[3]{\ensuremath{\langle #1 | #2 | #3 \rangle}}

\newcommand{\ie}{{\it i.e. }}

\begin{document}

\title{A quantitative theory of coherent delocalization}
\author{Federico Levi and Florian Mintert}
\address{Freiburg Insitute for Advanced Studies, Albert-Ludwigs University of Freiburg, Albertstr. 19, 79104 Freiburg, Germany.}

\begin{abstract}
We define a quantitative measure of coherent delocalization; similarly to the concept of entanglement measures, we require that a measure of coherent delocalization may never increase under processes that do not create coherent superpositions. 
After a complete characterization of such processes, we prove that a set of recently introduced functions that characterize coherent delocalization never grow under such processes and thus are indeed valid measures.

\end{abstract}


\maketitle

\section{Introduction}

Coherent transport is a phenomenon that occurs in various fields of physics.
Its fundamental properties are illustrated by the multi-slit experiment, where the maximum of the interference pattern is limited by the number of coherently illuminated slits.
Such an enhancement of arrival probability can be found in any system in which the propagating object can take several path alternatives coherently,
{\it i.e.} can be coherently delocalized.
Excitons in bio-molecular networks are just one example for such a situation,
but the identification of beatings in spectroscopic data obtained from natural \cite{Panitchayangkoon2011,Engel2007} and artificial \cite{Collini2009} samples which might be a signature of quantum coherence has substantially revived the interest in transport and interference.

 Any realistic system -- in particular a light harvesting complex -- is always subject to noise.
 Given the intrinsic stochastic nature of such processes, the propagating object would naturally take different paths depending on the current realisation of the noise.
Such a classical exploration of various paths, {\it i.e.} incoherent delocalization, however, does not result in any direct enhancement of the arrival probability.

Coherent delocalization gives rise to interference, which enhances transport efficiency if it is constructive \cite{Kramer1993,Scholak2011},
but may also reduce it if destructive \cite{Lattices1956,Plenio2008,Rebentrost2009c}.
Incoherent delocalization on the other hand does not have the potential for strong enhancement of transport efficiency,
but results in transport that is typically substantially more robust against perturbations than entirely coherent transport.
It is thus the interplay of coherent and incoherent delocalization that governs the efficiency with which the propagating object reaches a given final state.

Here, we introduce a quantitative concept of coherent delocalization that shall help to improve our understanding of the distinction between coherent and incoherent aspects of quantum transport.  A rigorous approach to distinguish these two aspects has been sought for a while \cite{Smyth2012}.
Originally, this was attempted in terms of time-averaged quantities, which could provide a rough estimation of delocalization  in specific systems \cite{Kuhn1997,Leegwater1996}. Recently, the transfer of instruments from entanglement theory resulted in a rather rigorous distinction between coherent and incoherent delocalization \cite{Caruso2010, Fassioli2010b},
and tools that verify a coherent delocalization of a given range (analogous to the number of coherently illuminated slits) are available \cite{Levi2013}.
So far, however, a clear quantitative meaning of such tools is unavailable,
but would be necessary to properly assess the potential for interference and resulting enhancement of transport. 
 
In the following, we consider a general setting to describe quantum transport. The formal framework that we introduce is not limited to a specific physical scenario; it is applicable for example to the case of multi-armed interferometers, excitation transport in bio-molecular networks \cite{Scholes2011}, electron transport in a network of quantum dots \cite{Greentree2004}, photons in an optical network \cite{Schreiber2010} or more generally to assess coherence in quantum random walks. Irrespective of the specific physical system, the propagating object can adopt a given number of states $\ket{\Psi_i}$: those may, for example, indicate which path is taken by a particle in an interferometer, which individual chromophore is excited in a light harvesting complex, or they may indicate which excitonic eigenstate of the same system is populated. Once this set is determined, one may strive for the question over how many of these states the object is delocalized and to what extent the delocalization is coherent.

Given that coherent delocalization is a rather abstract concept one needs a clear notion to define its quantification.
This is similar to the theory of entanglement, where a quantitative notion has been found in terms of evolutions (commonly referred to as quantum channels) that may not create entanglement. The central requirement for a measure of entanglement is that it may never increase under these channels \cite{Bennett1996a}.
We follow here the same line of arguments: we identify channels that can not induce a coherent superposition of the states $\ket{\Psi_i}$, which we refer to as incoherent channels. Once this is established, our central requirement for a quantity to describe coherent delocalization over a given range of states quantitatively is that is may not grow under incoherent channels. Such concept is well established in the theory of entanglement and it starts to be increasingly employed also in the context of quantum coherence, as independently done in \cite{Baumgratz2013}.

\section{Coherent delocalization}

When the propagating object is isolated from its external environment, {\it i.e.} in absence of any noise, it is described by a pure state $\ket{\chi}=\sum_{i=1}^n \xi_i \ket{\Psi_i}$, where $n$ is the dimension of the system. The amplitudes $\xi_i$ give the probabilities $|\xi_i|^2$ to find the object in a specific state of the set $\ket{\Psi_i}$.
If only one amplitude $\xi_j$ is non-vanishing, then the object is completely localized on the state $\ket{\Psi_j}$.
A state $\ket{\chi}$ is defined coherently delocalized over $k$ states $\ket{\Psi_i}$, or in short $k$-coherent, if $k$ of the amplitudes $\xi_i$ are non-vanishing.

In the presence of noise, the correct instrument to describe the propagating object is a density matrix $\varrho=\sum_j p_j \prj{\chi_j}$.
It describes both quantum interference due to coherent delocalization in the pure state components $\ket{\chi_j}$ and classical averaging in terms of the probabilities $p_j$.
This interplay results in the fact that constructive interference originating from one pure state component may be compensated in this averaging process by destructive interference from other components.
That is, the potential of the state $\varrho$ to give rise to interference phenomena cannot be inferred exclusively from the coherence properties of the pure state components $\ket{\chi_j}$.
In particular, the description of the density matrix $\varrho$ in terms of the probabilities $p_j$ and states $\ket{\chi_j}$ is not unique. The eigensystem provides one decomposition, but there is a continuous set of decompositions with typically non-orthogonal states $\ket{\chi_j}$ \cite{Schrodinger2008}.
If there is a decomposition of the density matrix composed of at most $k$-coherent states, than $\varrho$ describes a situation that can be accounted for by interference up to $k$ amplitudes and classical averaging. A mixed state $\varrho$ should therefore be considered $k$-coherent if there is no set of probabilities $p_j$ and at most ($k$-1)-coherent states $\ket{\chi_j}$ that are compatible with $\varrho$, {\it i.e.} that satisfy $\varrho=\sum_jp_j\prj{\chi_j}$.

\section{Measures for coherent delocalization}\label{section3}

Our objective is to provide a quantification of coherent delocalization rather than the mere qualitative distinction between different types of $k$-coherence defined above. As discussed in the introduction, the main requirement for such a quantification is that it may not grow under incoherent channels.
We formalize this concept later-on in section \ref{sec:incohproc},
but anticipate here that channels of this kind are the result of an average over two types of elementary incoherent processes.
The first is the modification of phase coherence, and it is described by operators of the form
\be
A_\ell=u_1\prj{\Psi_\ell}+u_2\sum_{j\neq \ell}\prj{\Psi_j}.
\label{eq:dephasing}
\ee
An average over this type of operations induces dephasing, \ie loss of phase coherence. The second type of elementary process is the incoherent hopping from $\ket{\Psi_\ell}$ to $\ket{\Psi_j}$, represented by the operators
\be
B_{j\ell}=\ket{\Psi_j}\bra{\Psi_\ell}.
\label{eq:hopping}
\ee
Any channel that can be described as an average over products of these two types of operators has to be considered incoherent, and a proper measure of coherent delocalization must never increase under such a channel.

Now that we have stated which condition to fulfil, we are in position to introduce the functions whose validity as quantifiers we prove later-on in section \ref{sec:monotonicity}.

Coherent delocalization first of all requires delocalization, \ie more than one finite diagonal matrix element $\matel{\Psi_i}{\varrho}{\Psi_i}$.
The larger (in terms of absolute values) the off-diagonal matrix elements are, the more pronounced is the coherent character of this delocalization.
One would thus expect that coherent delocalization could be characterized by the relation between diagonal and off-diagonal matrix-elements.
As shown in \cite{Levi2013} the quantities
\be
\label{hierarchy}
\tau_{kn}^{\varphi}(\varrho)=\big|\matel{\varphi_1}{\varrho}{\varphi_2}\big| -a_{kn}\sum_{j=1}^n\sqrt{\matel{\varphi_1^{(j)}}{\varrho}{\varphi_1^{(j)}}\matel{\varphi_2^{(j)}}{\varrho}{\varphi_2^{(j)}}}
\ee
with
\be
\label{eq:param}
a_{kn}=1/(n-k+1)\hspace{.2cm}\mbox{for}\hspace{.2cm}k\neq 2\hspace{.4cm}\mbox{and}\hspace{.4cm}a_{2n}=1/n,
\ee
are non-positive if $\varrho$ is not at least $k$-coherent, \ie coherently delocalized over $k$ of the states $\ket{\Psi_i}$. The states $\ket{\varphi_1}$ and $\ket{\varphi_2}$ are defined as
\begin{eqnarray}
\label{states}
\begin{aligned}
&\ket{\varphi_1}= C_1 \sum_{i=1}^n\frac{\beta_i}{\alpha_i}|\Psi_i\rangle\\
& \ket{\varphi_2}= C_2 \sum_{i=1}^n\frac{\beta_{i+n}}{\alpha_{i+n}}|\Psi_i\rangle
\end{aligned}
\end{eqnarray}
in terms of a set of $2n$ pairs of complex parameters $\{\alpha_i,\beta_i\}$, with $|\alpha_i|^2+|\beta_i|^2=1$.
The prefactors $C_i$ are given by $C_1=\prod_{i=1}^n \alpha_i$ and $C_2=\prod_{i=1}^n \alpha_{i+n}$.
The $n$ states $\ket{\varphi_1^{(j)}}$ and $\ket{\varphi_2^{(j)}}$ are obtained from $\ket{\varphi_1}$ and $\ket{\varphi_2}$
by a simple exchange of $\alpha_j,\beta_j$ and $\alpha_{j+n},\beta_{j+n}$, {\it i.e.} 
\bqa
\label{states2}
\begin{aligned}
\ket{\varphi_1^{(j)}}&=\ket{\varphi_1}\big|_{\alpha_j=\alpha_{j+n},\beta_j=\beta_{j+n}}\\
\ket{\varphi_2^{(j)}}&=\ket{\varphi_2}\big|_{\alpha_{j+n}=\alpha_{j},\beta_{j+n}=\beta_j}.
\end{aligned}
\eqa
The functionality of the definitions from \eref{hierarchy} to \eref{states2} can be understood in terms of pure states $\varrho=\prj{\chi}$. If $\xi_j=0$, the $j$-th term in the sum equals $\big|\matel{\varphi_1}{\chi\rangle\langle\chi}{\varphi_2}\big|$.
For a $k$-coherent state in an $n$-dimensional system there are $n-k$ such summands.
The prefactor $a_{kn}=1/(n-(k-1))$ thus makes sure that $\tau_{kn}^\varphi$ is non-positive for all states with less than $k$-coherence;
 for $k=2$ the prefactor can be chosen even smaller because in the case of complete localization each term in the sum equals $\big|\matel{\varphi_1}{\chi\rangle\langle\chi}{\varphi_2}\big|$.
 With these prefactors, \eref{hierarchy} is thus constructed to be non-positive for states that are less than $k$-coherent, and,
as shown in the appendix, for any at least $k$-coherent pure state $\ket{\chi}$ there exist vectors $\ket{\varphi_1}$ and $\ket{\varphi_2}$ which make $\tau_{kn}^\varphi(\prj{\chi})>0$. Finally, the functions defined in \eref{hierarchy} are convex as shown in \cite{Huber2010}; consequently, $\tau_{kn}^\varphi(\varrho)$ cannot be positive if there is a decomposition of the mixed state $\varrho$ into at most ($k-1$)-coherent states, \ie if $\varrho$ is less than $k$-coherent.
 
The potential of these functions to characterize coherent delocalization and its relation to transport efficiency was shown in \cite{Witt2013a}.
As we demonstrate in the following, each of the non-negative functions
\be
\label{deftaumax}
\mathcal{T}_{kn}(\varrho)=\max_{\varphi}\tau_{kn}^{\varphi}(\varrho)
\ee
is indeed non-increasing under incoherent operations,
and is thus not just a qualitative, but indeed a valid quantitative description of $k$-coherence.

\section{The formal framework}

As pointed out before, the following formalism is suited to describe several different physical scenarios which feature coherent transport. In order to keep the discussion simple, however, we choose a general terminology in terms of excitations and units. 
With excitation we indicate the to-be-transported entity, which could {\it e.g.} be an electronic excitation or a photon.
This excitation may be delocalized over a set of $n$ states $\ket{\Psi_i}$ ($i=1,...,n$).
Typically, the state $\ket{\Psi_i}$ denotes that the excitation is carried by the $i$-th out of $n$ physical entities that we refer to as units;
in the case of electronic excitations this could be chromophores; or they may be $n$ modes of the electromagnetic field in the case of photons in an interferometer. Regardless of the specific system, the units can be modeled as two levels systems.
These $n$ physical entities then define a composite system with $n$ components, so that each state $\ket{\Psi_i}$ can also be identified with an $n$-unit state
$\ket{\underbrace{0... 0}_{i-1}1\underbrace{0... 0}_{n-i}}$, where the $i$-th unit is excited. In this way the set of states $\ket{\Psi_i}$ is identifiable with the single excitation subspace of the full space which describes the $n$ units. For this reason, the term ``delocalization over $k$ states" possesses the same meaning as ``delocalization over $k$ units", and we use them interchangeably. The concept of $k$-coherence is then formally equivalent to the concept of $k$-partite entanglement in states that carry exactly one excitation \cite{Tiersch2012}.
This identification is going to be helpful in the following since it eases the analysis substantially. 
Since, however, the description in terms of single units is not necessary to define the concept of coherent delocalization,
our results also apply {\it e.g.} to the case where the states $\ket{\Psi_i}$ are excitonic eigenstates, and a clear physical identification in terms of an $n$-unit state is not necessarily available.

The states $\ket{\varphi_1}$, $\ket{\varphi_2}$, $\ket{\varphi_1^{(j)}}$ and $\ket{\varphi_2^{(j)}}$ as defined in \eref{states} and \eref{states2} can also be obtained from the projection of the the $n$-unit states on the subspace with a single excitation.
Originally, $\tau_{kn}$ is defined \cite{Levi2013} in terms of a $2n$-unit product vector $\ket{\Phi}=\ket{\Phi_1}\otimes\ket{\Phi_2}$ given by
\bqa
\begin{aligned}
\label{states3}
\ket{\Phi_1}&=\ket{\psi_1}\otimes\ket{\psi_2}\otimes...\otimes\ket{\psi_n},\\
\ket{\Phi_2}&=\ket{\phi_1}\otimes\ket{\phi_2}\otimes...\otimes\ket{\phi_n},
\end{aligned}
\eqa
where $\ket{\psi_i}$ and $\ket{\phi_i}$ are states for the $i$-th individual unit. The states $\ket{\Phi_1^{(j)}}$ and $\ket{\Phi_2^{(j)}}$, which enter the definition of $\tau^{\ket{\Phi}}_{kn}$ analogously to $\ket{\varphi_1^{(j)}}$ in \eref{deftaumax}, are obtained from $\ket{\Phi_1}$ and $\ket{\Phi_2}$ through an exchange of their $j$-th factor.
With the parametrization $\ket{\psi_j}=\alpha_j\ket{0}+\beta_j\ket{1}$ and  $\ket{\phi_j}=\alpha_{j+n}\ket{0}+\beta_{j+n}\ket{1}$,
one obtains the states defined above in \eref{states} and \eref{states2} through projection onto the single excitation subspace.

In the following it is convenient to work in the full space of the $n$ units: since we require that $\varrho$ carries exactly one excitation, we can switch between the definition of $\tau_{kn}^{\ket{\Phi}}$ in the full space and its restriction to single excitation subspace $\tau_{kn}^\varphi$ as we like. For this reason, we are going to use the same symbol $\mathcal{T}_{kn}$ for the maximum of $\tau_{kn}^\varphi$ and of $\tau_{kn}^{\ket{\Phi}}$.

\subsection{Incoherent channels}\label{sec:incohproc}

One central advantage of the description in terms of  the states of the individual units is that it permits the characterization of the operations that we anticipated as incoherent rather easily. 

A real physical evolution is represented formally by a classical stochastic average over single processes, each of which is described by an operator $F_i$. The set of all these operators $F_i$, called the Kraus operators, connects the initial state $\varrho$ to the final state $\varrho_F$ via a channel \cite{Nielsen2010}: 
\be
\label{eq:channel}
\varrho_F=\sum_iF_i\varrho F_i^\dagger,
\ee
where the additional constraint $\sum_iF_i^\dagger F_i=\um$ ensures the conservation of trace. 

An incoherent quantum channel is composed by incoherent Kraus operators, \ie is an average over incoherent processes. Since none of these should coherently delocalize an initially localized excitation, any operator $F_i$ must be {\it local}, which means that it is given by the tensor product of operators acting on individual units, {\it i.e.}
\be
F_i=f_i^{(1)}\otimes...\otimes f_i^{(n)}.
\label{eq:localoperator}
\ee
In addition, we require that no such operator may change the number of excitations.

Each single-unit operator $f_i^{(j)}$ can be expanded in an operator basis.
Rather than using the typically employed Pauli matrices, we use in the following the raising operator $\sigma_+=(\sigma_x+i\sigma_y)/2$, the lowering operator $\sigma_-=(\sigma_x-i\sigma_y)/2$,
the regular Pauli matrix $\sigma_z$ and the identity $\um$.
Any single-unit operator $\sigma_0=b\mathbbm{1}+d\sigma_z$ with complex coefficients $b$ and $d$ does not modify the number of excitations; $\sigma_+$ and $\sigma_-$ create and annihilate an excitation respectively.

We show in the following that each single-unit operator $f_i^{(j)}$ is either of the form $\sigma_0$, $\sigma_+$ or $\sigma_-$.
To this end, we consider a general form of $F$\footnote{the index $i$ is suppressed, as the following holds for all operators $F_i$ irrespective of their specific label}, not necessarily local, and expand it in the above introduced basis for the first subsystem. This gives
\begin{equation}
\label{excons}
F=c_0 (\sigma_0\otimes\mathcal{A}_0)+c_+(\sigma_+\otimes \mathcal{A}_-)+c_-(\sigma_-\otimes \mathcal{A}_+).
\end{equation}
$F$ conserves the number of excitations exactly if $\mathcal{A}_-$ and $\mathcal{A}_+$ annihilate and create an excitation respectively, and $\mathcal{A}_0$ conserves the number of excitations.
The operators $\sigma_0$, $\sigma_+$ and $\sigma_-$ are mutually orthogonal, and so are the operators $\mathcal{A}_0$, $\mathcal{A}_-$ and $\mathcal{A}_+$, which act on the remaining $n$-1 units.
$F$ is thus in the shape of a Schmidt decomposition \cite{Horodecki2009}, and one can directly conclude that it is of product form exactly if two of the three coefficients $c_0$, $c_+$ and $c_-$ vanish.
In a similar fashion one can proceed to investigate the locality of the operators $\mathcal{A}_0$, $\mathcal{A}_-$ and $\mathcal{A}_+$. By induction, one obtains that any operator in \eref{eq:localoperator} that conserves the number of excitations contains only operators of the form $\sigma_0$, $\sigma_+$ and $\sigma_-$ as factors; linear combinations would not be excitation conserving or would be nonlocal.

Restricted to the single excitation subspace discussed at the beginning of this section, the operator acting on $n$ units constructed from the single-unit operator $\um(u_1+u_2)/2+\sigma_z^{(\ell)}(u_1-u_2)/2$ coincides exactly with $A_\ell$ defined in \eref{eq:dephasing}, while $B_{\ell j}$ originates from the joint contribution of the two single unit operators $\sigma_+^{(j)}\otimes\sigma_-^{(\ell)}$. We have therefore proven that indeed the only two elementary processes which do not create coherence and conserve the number of excitation are local dephasing and excitation hopping. All local excitation conserving Kraus operators acting on $n$ units, which we shall from now on compactly refer to as incoherent Kraus operators, can be obtained in terms of these two processes.

\subsection{A measure for coherent delocalisation decreases under incoherent channels}
\label{sec:monotonicity}

Our goal is to prove that $\mathcal{T}_{kn}(\varrho_F)\le\mathcal{T}_{kn}(\varrho)$ for all channels as defined in \eref{eq:channel} which are composed exclusively  by incoherent Kraus operators.

The functions $\tau_{kn}^{\ket{\Phi}}$ are convex, as shown in \cite{Huber2010}. This allows us to conclude that
\be
\mathcal{T}_{kn}(\sum_iF_i\varrho F_i^\dagger)\le\max_{\Phi}\sum_i\tau_{kn}^{\ket{\Phi}}(F_i\varrho F_i^\dagger).
\label{eq:convex}
\ee
From the definition of $\tau_{kn}$ in \eref{hierarchy} it follows that
\be
\tau_{kn}^{\ket{\Phi}}(F_i\varrho F_i^\dagger)=\tau_{kn}^{F_i^\dagger\otimes F_i^\dagger\ket{\Phi}}(\varrho).
\ee
The state $F_i^\dagger\otimes F_i^\dagger\ket{\Phi}$ is still of product form because the Kraus operators $F_i$ are local as defined in \eref{eq:localoperator}.
This state is however in general not normalized. It is thus convenient to introduce the renormalized state
\be
\ket{\tilde{\Phi}_i}=\frac{F_i^\dagger\otimes F_i^\dagger\ket{\Phi}}{\sqrt{\matel{\Phi}{F_iF_i^\dagger\otimes F_iF_i^\dagger}{\Phi}}}.
\ee
Since $\tau_{kn}$ is a homogeneous function, we have
\be
\label{eq:rescale}
\tau_{kn}^{F_i^\dagger\otimes F_i^\dagger\ket{\Phi}}(\varrho)=\tau_{kn}^{\ket{\tilde\Phi_i}}(\varrho)\ \underbrace{\sqrt{\matel{\Phi}{F_iF_i^\dagger\otimes F_iF_i^\dagger}{\Phi}}}_{\eta_i}.
\ee
The maximization in the definition of ${\cal T}_{kn}$ runs over all product vectors,
and since the state vectors $\ket{\tilde\Phi_i}$ are in product form too,
a maximization over the state vectors $\ket{\tilde\Phi_i}$ can never yield something larger than maximization over $\ket{\Phi}$, which implies that $\tau_{kn}^{\ket{\tilde{\Phi}_i}}(\varrho)\leq\mathrm{max}_{\tilde{\Phi}_i}\tau_{kn}^{\ket{\tilde{\Phi}_i}}(\varrho)\leq\mathcal{T}_{kn}(\varrho)$. That is, together with \eref{eq:convex}, we can conclude that
\bqa
\begin{aligned}
\mathcal{T}_{kn}(\sum_iF_i\varrho F_i^\dagger)&\le\max_{\Phi}\sum_i\tau_{kn}^{\ket{\tilde\Phi_i}}(\varrho)\ \eta_i\\
&\le\mathcal{T}_{kn}(\varrho)\ \max_\Phi\sum_i\eta_i.\label{a4}
\end{aligned}
\eqa
With the Cauchy-Schwartz inequality $\big|\sum_ix_iy^*_i\big|\le\sqrt{\sum_i|x_i|^2}\sqrt{\sum_i|y_i|^2}$, the right hand side can conveniently be bounded from above:
\numparts
\bqa\label{factor}
\max_{\Phi}\sum_i\eta_i&=&\max_{\Phi_1 \Phi_2}\sum_i\sqrt{\matel{\Phi_1}{F_iF_i^\dagger}{\Phi_1}\matel{\Phi_2}{F_iF_i^\dagger}{\Phi_2}}\label{eq:app1}\\
&\le&\max_{\Phi_1 \Phi_2}\sqrt{\matel{\Phi_1}{\sum_iF_iF_i^\dagger}{\Phi_1}}\sqrt{\matel{\Phi_2}{\sum_iF_iF_i^\dagger}{\Phi_2}}\label{eq:app2}.
\eqa
\endnumparts
The two factors in \eref{eq:app2} are positive; the maximum of the product is therefore obtained as the product of the individual maxima. Since these maxima coincide, we have:
\bqa
\label{eq:app3}
\begin{aligned}
\max_{\Phi}\sum_i\eta_i&\le\left(\max_{\Phi_1}\sqrt{\matel{\Phi_1}{\sum_iF_iF_i^\dagger}{\Phi_1}}\right)^2=\\
&=\max_{\Phi_1}\,\matel{\Phi_1}{\sum_iF_iF_i^\dagger}{\Phi_1}\label{eq:app4}\ .
\end{aligned}
\eqa
All together, we arrive at
\begin{equation}
\label{tausep}
\mathcal{T}_{kn}\left(\sum_iF_i\varrho F_i^\dagger\right)\leq\mathcal{T}_{kn}(\varrho)\,\max_{\Phi_1}\left(\matel{\Phi_1}{\sum_iF_iF_i^\dagger}{\Phi_1}\right).
\end{equation}
The right hand side of \eref{tausep} involves the expression $\sum_iF_iF_i^\dagger$,
and conservation of trace implies $\sum_iF_i^\dagger F_i=\um$.
If the condition $\sum_iF_iF_i^\dagger=\um$ was satisfied, then the proof would be complete,
but, at this stage, we can only assert that
$\mathcal{T}_{kn}$ is non-increasing under any incoherent channel that satisfies $\sum_i [F_i,F^\dagger_i]=0$. 

Starting from \eref{tausep} it takes only a few minor steps to finish the proof.
First of all, one may observe that $\sigma_0$ is normal, {\it i.e.} $[\sigma_0,\sigma_0^\dagger]=0$; this means that $\mathcal{T}_{kn}$ is non-increasing under dephasing processes. Incoherent hopping processes, on the other hand, are not normal, {\it i.e.} $\sigma_+^{(j)}\otimes\sigma_-^{(\ell)}$ does not commute with its adjoint
$(\sigma_+^{(j)}\otimes\sigma_-^{(\ell)})^\dagger=\sigma_-^{(j)}\otimes\sigma_+^{(\ell)}$,
and $\sum_i [F_i,F^\dagger_i]\neq 0$ for a general incoherent channel.
It is however possible to assert that $(\sigma_+^{(j)}\otimes\sigma_-^{(\ell)})\varrho(\sigma_-^{(j)}\otimes\sigma_+^{(\ell)})$ describes a situation of perfect localization on the state $\ket{\Psi_j}$,
because there is only a single excitation.
That is $\mathcal{\tau}^{\ket{\Phi}}_{kn}(F_i\varrho F_i^\dagger)\le 0$ for any product vector $\ket{\Phi}$ if $F_i$ contains a term $\sigma_+^{(j)}\otimes\sigma_-^{(\ell)}$.

Including all these points, one finally obtains
\numparts
\bqa
\mathcal{T}_{kn}(\varrho_F)&=&\mathcal{T}_{kn}(\sum_iF_i\varrho F_i^\dagger)\label{eq:proof1}\\
&\le&\mathcal{T}_{kn}\Bigl(\sum_{i|dp}F_i\varrho F_i^\dagger\Bigr)+\mathcal{T}_{kn}\Bigl(\sum_{i|h}F_i\varrho F_i^\dagger\Bigr)\label{eq:proof2}\\
&=&\mathcal{T}_{kn}\Bigl(\sum_{i|dp}F_i\varrho F_i^\dagger\Bigr).\label{eq:proof3}
\eqa
\endnumparts
From \eref{eq:proof1} to \eref{eq:proof2} we divided the sum into the sum ($\sum_{i|dp}$) over Kraus operators that describe pure dephasing and a sum ($\sum_{i|h}$) that contains only Kraus operators that include hopping terms.
The inequality holds due to convexity of $\mathcal{T}_{kn}$ which is inherited from the convexity of $\tau^{\ket{\Phi}}_{kn}$.
Due to convexity also the second term in \eref{eq:proof2} vanishes, since $\mathcal{T}_{kn}(F_i\varrho F_i^\dagger)$ vanishes for any Kraus operator that includes hopping because $\mathcal{T}_{kn}$ is non-negative \footnote{The choice $\beta_i=0$ $\forall i$ yieds in fact a value of $\tau^\varphi_{kn}(\varrho)=0$ for any state $\varrho$ within the single-excitation subspace, so that $\max_{\varphi}\tau^\varphi_{kn}(\varrho)\geq 0$. }.
Using \eref{tausep}, we finally obtain
\be
\mathcal{T}_{kn}(\varrho_F)\le\mathcal{T}_{kn}(\varrho)\max_{\Phi_1}\left(\matel{\Phi_1}{\sum_{i|dp}F_i F_i^\dagger}{\Phi_1}\right)\label{eq:proof4}.
\ee
All Kraus operators in \eref{eq:proof4} commute with their adjoint, so that
$\sum_{i|dp}F_iF_i^\dagger=\sum_{i|dp}F_i^\dagger F_i$.
Conservation of trace implies
$\sum_{i|dp}F_i^\dagger F_i+\sum_{i|h}F_i^\dagger F_i=\um$,
and since $\sum_{i|h}F_i^\dagger F_i$ is a positive operator,
the operator inequality $\sum_{i|dp}F_i^\dagger F_i\le\um$ holds.
That is, no expectation value of $\sum_{i|dp}F_i^\dagger F_i$ exceeds the value of unity; we thus arrive at the desired result
\be
\mathcal{T}_{kn}(\varrho_F)\le\mathcal{T}_{kn}(\varrho),
\ee
which rigorously verifies that $\mathcal{T}_{kn}$ can never increase under incoherent dynamics.

\section{Illustration}

In order to illustrate the concept of incoherent channels and their distinction from coherent dynamics we discuss a few examples based on commonly employed models. The optimization required for the assessment of \eref{deftaumax} can be performed with standard routines, as included in current computer algebra packages. The linear scaling of the to-be-optimized parameters with the system size permits to treat considerable dimensions. Potential problems with local maxima can be avoided rather reliably by repeating the optimization routine with different initial conditions that are distributed over the whole parameter space.

The typical Hamiltonian for the description of coherent excitation transport reads
\be
{\cal H}=\sum_{i\neq j}^n\lambda_{ij}(\sigma_+^{(i)}\otimes\sigma_-^{(j)}+\sigma_-^{(i)}\otimes\sigma_+^{(j)})\ ,
\ee
where $\lambda_{ij}$ is the coupling strength between the $i$-th and $j$-th unit. Such a Hamiltonian models a system where the excitation travels due to coherent interaction. The propagator ${\cal U}(t)=e^{-i{\cal H}t}$ induced by this Hamiltonian conserves the number of excitations, but it is {\em not} of product form, \ie not incoherent. This can be seen explicitly in the exemplary case of $n=2$, where it reads
\be
\label{eq:ham}
{\cal U}(t)=\cos(\lambda_{12}t)\mathcal{P}_s+i\sin(\lambda_{12}t)(\sigma_+\otimes\sigma_- + \sigma_-\otimes\sigma_+),
\ee
in terms of the projector $\mathcal{P}_s$ on the single excitation subspace $\mathcal{P}_s=\prj{0}\otimes\prj{1}+\prj{1}\otimes\prj{0}$ and the creation and annihilation operators $\sigma_+$ and $\sigma_-$. $\mathcal{H}$ thus induces a dynamics which may create or enhance coherent delocalization.

Such a coherent dynamics needs to be distinguished from incoherent hopping, which might for example be described by a Master equation $\frac{d}{dt}\varrho(t)=\mathcal{D}(\varrho(t))$ with
\bqa
\label{eq:lindblad}
\begin{aligned}
\mathcal{D}(\varrho(t))&=\sum_{i>j}^n\mathcal{D}_{ij}(\varrho(t))\\
\mathcal{D}_{ij}(\varrho(t))&=\sum_{k=1}^4\gamma_{ij}\left(G_k^{(ij)}\varrho(t) \,G_{k}^{\dagger(ij)}-\frac{1}{2}\{G_{k}^{\dagger(ij)} G^{(ij)}_{k},\varrho(t)\}\right),
\end{aligned}
\eqa
where the Lindblad operators $G_k^{(ij)}$ are given by
\bqa
\begin{aligned}
G_{1}^{(ij)}&=\sigma^{(i)}_+\otimes\sigma^{(j)}_-, \qquad G_{2}^{(ij)}=\sigma^{(i)}_-\otimes\sigma_+^{(j)},\\
G_3^{(ij)}&=(P_1^{(i)}\otimes P_1^{(j)}-P_0^{(i)}\otimes P_0^{(j)})/\sqrt{2},\\
G_4^{(ij)}&=\sigma^{(i)}_z\otimes\sigma^{(j)}_z/(2\sqrt{2}).
\end{aligned}
\eqa
The operators $P_{1(0)}^{(i)}$ are the projectors on the excited (ground) state of the $i$-th unit.
The solution to a master equation can be expressed in terms of a set of Kraus operators which define a quantum channel $\varrho(t)=\sum_i F_i(t)\varrho(0)F^\dagger_i(t)$, as discussed in section \ref{sec:incohproc}.
In the specific case given by \eref{eq:lindblad}, for $n=2$ these Kraus operators read (indices have been omitted, since there are only two units)
\be
\begin{split}
\label{eq:kraus2n}
F_1&=\sqrt{\frac{1-\Gamma_t^2}{2}}\sigma_+\otimes\sigma_-,\\
F_3&=\sqrt{1-\Gamma_t}P_1\otimes P_1, \\
F_5&=\frac{1-\Gamma_t}{\sqrt{2}}P_1\otimes P_0,\\
F_7&=\sqrt{\Gamma_t}\um\otimes\um,
\end{split}
\qquad
\begin{split}
F_2&=\sqrt{\frac{1-\Gamma_t^2}{2}}\sigma_-\otimes\sigma_+,\\
F_4&=\sqrt{1-\Gamma_t}P_0\otimes P_0,\\
F_6&=\frac{1-\Gamma_t}{\sqrt{2}}P_0\otimes P_1,\\
\end{split}
\ee
with $\Gamma_t=\exp(-\gamma_{12} t)$. All the $F_i$ conserve the number of excitations and are local: they therefore describe a purely incoherent dynamics. Indeed, in the single excitation subspace, $F_1$ and $F_2$ are of the form of $B_{j\ell}$ as defined in \eref{eq:dephasing} and $F_3$ to $F_7$ are of the form $A_\ell$ as defined in \eref{eq:hopping}.

\begin{figure}
\centering
\includegraphics[width=0.7\textwidth]{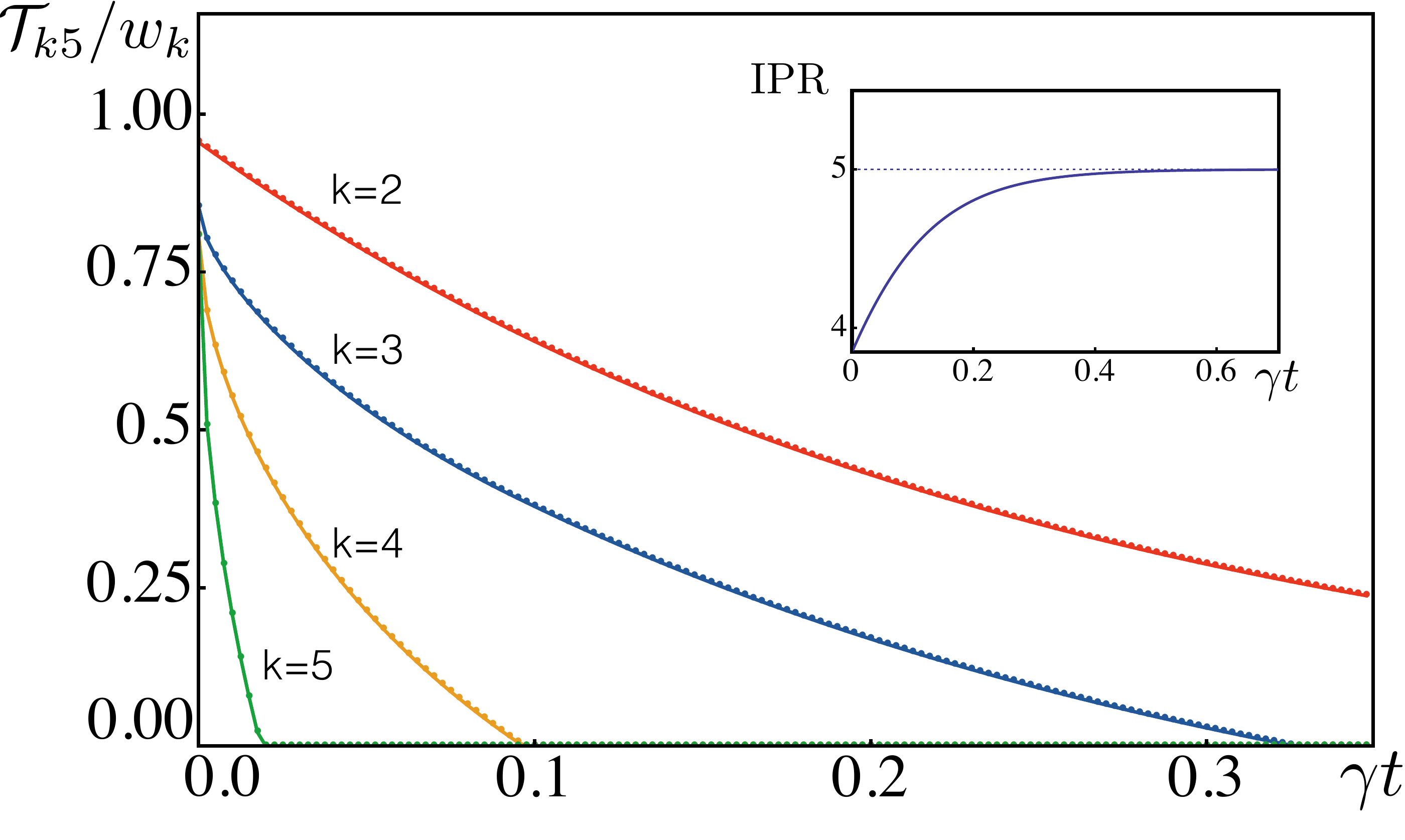}
\caption{
Coherent delocalization characterized by $\mathcal{T}_{kn}/w_k$ with the normalization constant $w_k=\mathcal{T}_{k5}(\ket{W})$, where $\ket{W}$ is the state with maximal coherent delocalization over 5 units. The time evolution for the initial state $\sqrt{\frac{1}{10}}(\ket{1}+\ket{5})+\sqrt{\frac{2}{10}}(\ket{2}+\ket{4})+\sqrt{\frac{4}{10}}\ket{3}$ for a system with five units is induced by the master equation defined by \eref{eq:lindblad}, with $\gamma_{ij}=\gamma$. As expected all $\mathcal{T}_{k5}$ decrease motononically despite the increase of delocalization as shown by the IPR in the inset.} 
\label{fig:incoherentcase}
\end{figure}

Having verified that \eref{eq:lindblad} induces incoherent dynamics for $n=2$ allows us to draw this conclusion also for $n>2$ with help of the Trotter expansion \cite{Trotter1959}: since the solution $\varrho(t)=e^{{\cal D}t}\varrho(0)$ can be expressed in terms of solutions of the two-site system via $e^{{\cal D}t}=\lim_{m\to\infty}\left(\Pi_{ij}e^{{\cal D}_{ij}t/m}\right)^m$,
the dynamics induced by \eref{eq:lindblad} can be given in terms of incoherent Kraus operators for any system size $n$. Consequently, coherent delocalization can not grow under this dynamics.
This is exemplified in figure~\ref{fig:incoherentcase}, which shows the behaviour of $\mathcal{T}_{kn}(\varrho)$ for $n=5$ and $k$ ranging from $2$ to $5$ for a time evolution induced by \eref{eq:lindblad}. Initially, the excitation is delocalized coherently but it is not maximally delocalized.
The dynamics then induces an increase of delocalization, which is indicated by the growth of the inverse participation ratio, defined as $\mathrm{IPR}(t)=1/\sum_iq^2_i(t)$, where the population of the $i$-th unit is given by $q_i(t)=\mathrm{Tr}(\varrho(t)\prj{\Psi_i})$ \cite{Thouless1974}. The behaviour of the IPR is shown in the inset of figure~\ref{fig:incoherentcase}.
Since the dynamics is incoherent, however, {\em coherent} delocalization may not increase as correctly identified by $\mathcal{T}_{kn}$. In the stationary state, that is reached for $t\to\infty$, the excitation is delocalized completely over the entire system but this delocalization is completely incoherent, so that $\mathcal{T}_{kn}$ vanishes for all $k$. 

A qualitatively similar behavior can be observed in a system with a disordered Hamiltonian and local phase noise. For sufficiently large disorder, the eigenstates of the system Hamiltonian are strongly localized.
If the system is initially prepared in an eigenstate, the dephasing will result in a growing delocalization with an increasing IPR.
We found that the measures $\mathcal{T}_{kn}$, however, do not increase; that is they correctly assess the incoherent nature of this delocalization.

\begin{figure}
\centering
\includegraphics[width=0.7\textwidth]{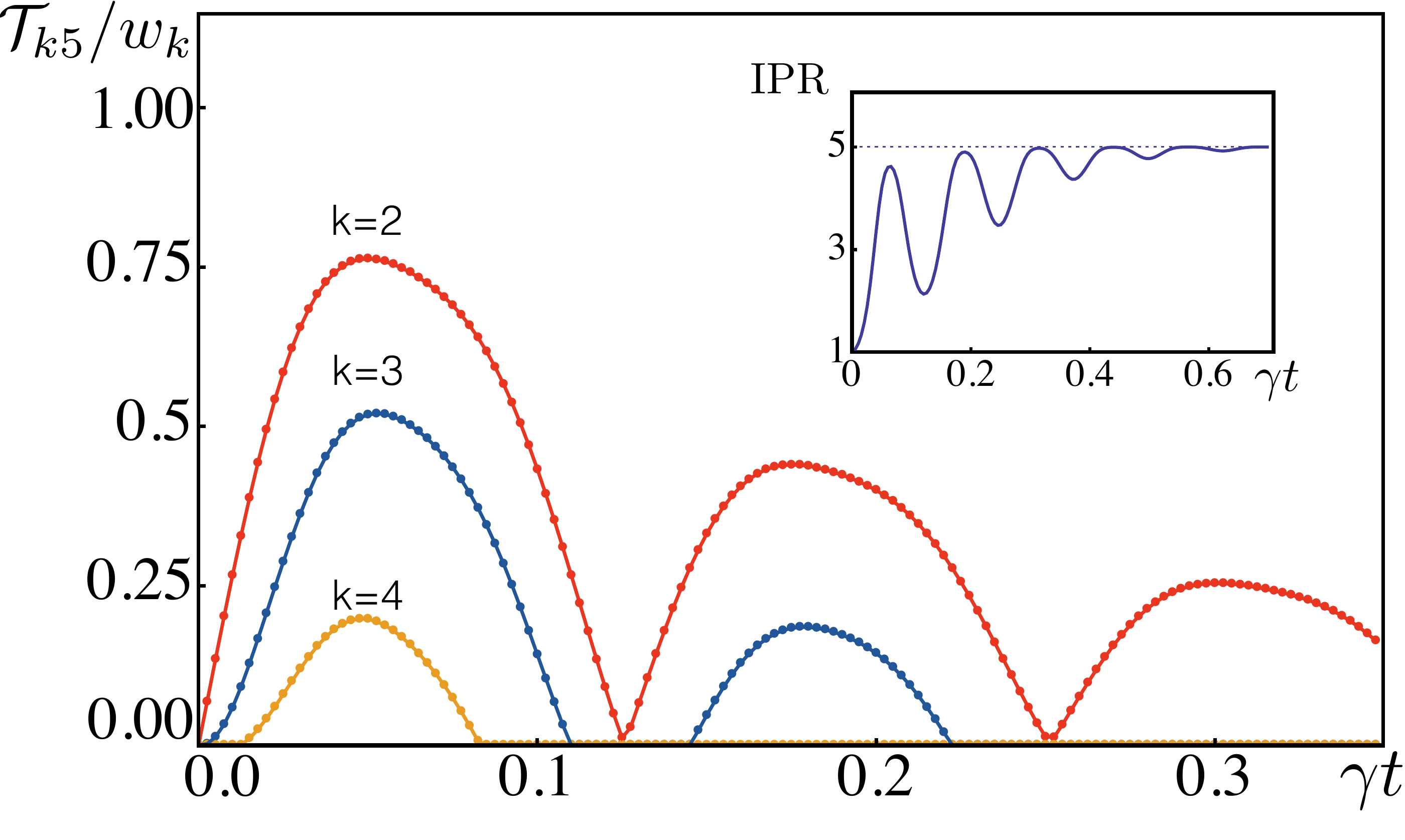}
\caption{
Coherent delocalization characterized by $\mathcal{T}_{kn}/w_k$ for an evolution with coherent and incoherent aspects defined respectively by \eref{eq:ham} and  \eref{eq:lindblad}  (see the caption of figure \ref{fig:incoherentcase} for the definition of $w_k$). The strength of the coherent and incoherent interactions are given by $\lambda_{ij}=10\gamma$. Initially the excitation is perfectly localized. Due to the influence of the coherent interaction, coherent delocalization over up to four units is created until $\gamma t\approx 0.05$. For longer times, the noise reduces the coherent character of the delocalization, but generates incoherent delocalization as one can see from the behaviour of the IPR in the inset.}
\label{fig:inbetweencase}
\end{figure}

In general, however, a coherent interaction can also increase the coherent delocalization of an excitation.
To exemplify this we consider a Master equation where in addition to the incoherent dynamics described by ${\cal D}$ \eref{eq:lindblad}, there is a coherent term described by the Hamiltonian $\mathcal{H}$ defined in \eref{eq:ham} with $\lambda_{ij}=\lambda=10\gamma$;  Figure ~\ref{fig:inbetweencase} depicts the behaviour of $\mathcal{T}_{kn}$ with $n=5$ after initialization with a perfectly localized excitation. Due to the coherent part of the dynamics, coherent delocalization initially grows. In the case of perfectly coherent dynamics one would observe strictly periodic motion, but, due to the additional incoherent character, one can observe an overall attenuation of $\mathcal{T}_{kn}$.
In particular, the incoherent contribution is so large that no coherent delocalization over five units can be verified.
Similarly to the completely incoherent dynamics above in figure~\ref{fig:incoherentcase}, also in this case the IPR asymptotically reaches its maximal value.

\section{Conclusions}

Recently, tremendous progress in both the experimental and the theoretical analysis of exciton transport in complex bio-molecular systems has been achieved.
On the experimental side, spectroscopy provides data with resolutions that were unthinkable a few years ago \cite{Collini2013}, while theoretical methods permit to simulate exciton dynamics very accurately despite firm environment coupling and strong non-Markovian dynamics \cite{Pachon2012}.
The improved available data, in turn, asks for reliable techniques that permit a rigorous interpretation.
For example, the question whether beating signals would permit to draw conclusion about quantum coherence has induced a very controversial debate so far \cite{Lambert2012}.
To some extent, this is due to the fact that we have a rather vague concept of quantum coherence, and a more solid theoretical footing is just about to be developed.

With the quantitative concept of quantum coherence presented here, we provide a rigorous basis for the analysis of coherence properties. 
At this stage, the present tools are applicable rather to theoretical treatments that permit to construct the complete density matrix.
Since, however, only a limited number of density matrix elements are required to access $\tau_{kn}$, a direct experimental observation seems conceivable in {\it e.g.} artificially designed \cite{Collini2009} or, given further improvement of spectroscopic techniques, even in actual light harvesting complexes.

Financial support by the European Research Council under the project Odycquent is gratefully acknowledged.

\appendix
\section{}

We show here that $\mathcal{T}_{kn}(\prj{\chi})$ is positive for any pure state $\ket{\chi}$ that is at least $k$-coherent. For this purpose we will construct a set of not necessarily normalized state vectors $\ket{\varphi_i}$ such that $\tau_{kn}^{\varphi}$ is positive. Since $\tau_{kn}^\varphi$ is a homogeneous function of the $\ket{\varphi_i}$, also the normalized state vectors will yield a positive value, which makes $\mathcal{T}_{kn}$ positive due to its definition as maximum of $\tau_{kn}^\varphi$.

For pure states \eref{hierarchy} reduces to
\be
\label{eq:hierarchypure}
\tau_{kn}^\varphi(\prj{\chi})=\underbrace{|\ovl{\chi}{\varphi_1}\ovl{\chi}{\varphi_2}|}_{A}-a_{kn}\sum_{j=1}^n\underbrace{|\ovl{\chi}{\varphi^{(j)}_1}\ovl{\chi}{\varphi_2^{(j)}}|}_{B_j},
\ee
where we have introduced the short-hand notations $A$ and $B_j$.

In the following, we consider a $k$-coherent state $\ket{\chi}=\sum_{\ell=1}^n \xi_\ell \ket{\Psi_\ell}$; without loss of generality we assume that $\xi_\ell\neq0$ for $\ell=1,...,k$ and $\xi_\ell = 0$ for $i=k+1,..,n$. Since $B_j=A$ for $j=k+1,...n$, as explained in section \ref{section3}, \eref{eq:hierarchypure} reduces to
\be
\tau_{kn}^\varphi(\prj{\chi})=A\Big(1-a_{kn}(n-k)\Big)-\sum_{j=1}^k B_j.
\ee
Since we can resort to un-normalized states, we choose $\alpha_i=1$ for the coefficients defined in \eref{states} to obtain
\be
\label{eq:A}
A=\Big|\underbrace{\sum_{i=1}^k \beta_i\xi^*_i}_x\Big|\Big|\underbrace{\sum_{i=1}^k\beta_{i+n}\xi^*_i}_{y}\Big|,\quad \text{and}
\ee
\be
B_j=\Big|x +\xi^*_j(\beta_{j+n}-\beta_j)\Big|\Big| y + \xi^*_j(\beta_{j}-\beta_{j+n})\Big|.
\ee
We can pick values for the coefficients $\beta_i$ ($i=1,...,k$) such that $|x|>0$. Given that the state amplitudes $\xi_j$ are non-vanishing for $i=1,..,k,$ one may choose
\be
\label{eq:betachoice}
\beta_{j+n}=\beta_j-\frac{x}{\xi^*_j},
\ee
which leads to $B_j=0$ for $j=1,..,k.$ With this choice for the parameters one obtains
\be
y=\sum_{i=1}^k \beta_{i+n}\xi_i = (1-k)x,
\ee
which is non vanishing since $x\neq 0$ and $k\neq 1$. That is, all-together, we found a choice for the $\beta_i$ such that 
\be
\tau_{kn}=\left(1-a_{kn}(n-k)\right)|x||y|,
\ee
which is positive since $a_{kn}(n-k)<1$. This, in turn, proves that $\mathcal{T}_{kn}$ is positive for any $k$-coherent pure state. Since $\mathcal{T}_{kn}\leq \mathcal{T}_{k'n}$ for $k>k'$, $\mathcal{T}_{kn}$ is positive for any pure state that is at least $k$-coherent.

\bibliography{../../tex/Bib/referenzen,../Bib/libraryjabref,../Bib/library}

\end{document}